# Dancing with the Unexpected and Beyond

## The Use of AI Assistance in Design Fiction Creation

Yiying Wu

School of Design, The Hong Kong Polytechnic University, Hong Kong, China, bowyiying.wu@polyu.edu.hk

Yunye Yu

School of Foreign Languages, Southeast University, Nanjing, China, yunye.yu@seu.edu.cn

Pengcheng An

Southern University of Science and Technology, Shenzhen, China, anpc@sustech.edu.cn

The creation process of design fiction is going participatory and inclusive with non-experts. Recognizing the potential of artificial intelligence in creativity support, we explore the use of AI assistance in creating design fiction. This investigation is based on a workshop on 'future work in 2040' with Chinese youth (N=20). We look into fiction quality, participants' experiences with the AI agent, and their ways of incorporating those texts into writing. Our findings show that human writers while responding to messy and unexpected AI-generated texts, can elevate the richness and creativity in writing and initiate joyful and inspirational interactions. Furthermore, for the design of AI assistance in creativity support, we suggest two implications of enhancing interactional quality between human and AI and prompt programming. Our study indicates the potential of applying design fiction outside the design context using a more inclusive approach for future speculation with critical reflection on technology.

CCS CONCEPTS • Human-centred computing • Human-computer interaction • Empirical studies in HCI

**Keywords:** Design fiction, Human-AI interaction, Creative writing, Creativity support tools

### 1 INTRODUCTION

How will technological futures look? What will be the societal and ethical implications of disruptive technologies? These questions are often answered by experts such as engineers, developers, or technological futurists. However, the expert-based practice is criticised for the reproduction of the same hegemonic structure and the incapability of producing widen options of future [32], and the 'power imbalance' that disfavour certain groups and voices that eventually do not have a future [39, 51]. Opposing it, a participatory and inclusive approach to future speculation is increasingly seen as significant [54]. This approach is not only about the participatory process that should involve various voices but also refer to more plural and widening options the for future [10]. Studies show that laypeople have the future thinking and are especially good at contextualising macro impacts on their individual lives [24]. In practice, more non-experts are involved in creating future images and expressing preferences and opinions on technologies. In futures studies, people are invited to develop future scenarios in various themes of sustainable lifestyle [27], work [29], and energy [60].

Regarding future images, we are particularly interested in the provocative type of technological futures in which new technologies are not glorified and the complicated implications are discussed. By inviting people to create such scenarios, we wish to nurture the critical thinking of technology. This practice is resonated with the method of Design fiction from the HCI field. Design fiction has been a valuable tool in searching for the ethical and societal implications of emerging technology and developing critical thinking towards it. Despite the diversity in definitions, origins or approaches of design fiction practice [5, 41], the common elements refer to the narrative of the yet-to-exist with the 'discursive turn' [36]. Thus, we take the two elements of speculative and provocative from Design fiction to frame the future images to be created in our study. It is worthy of note that we are aware that one essential element of Design fiction is the making of 'diegetic prototype' [6] and the design fictions created from our case might not be considered 'designerly' enough for some designers. However, the primary aim of our work is to democratise the speculative and critical thinking to the larger mass. Thus, our study takes the narrative focus of Design fiction. It shares the affinity with other co-creation work of design fiction that is text-based [2, 19] and emphasizes the narrative potential in enquiries [18]. Also, the design fictions produced in this study would not focus on product development, but the development of narratives of new concepts and critical reflection.

The participatory thread in design fiction creation also leads to the challenge. The co-design literature claims that everyone is creative and the expert of their own experience [38, 52]. However, a challenge is how to facilitate people imagining and even debating the future that is yet-to-exist and new technologies that are unfamiliar? Taking up this challenge, we aim to facilitate and empower everyone in the creation and reflection. In the fields of design research and HCI, many researchers have developed tools and processes to involve non-designers' voices in innovating new technologies and commenting on future scenarios [37]. We will review the related work in detail in the later section. Here, this paper looks into the potential of artificial intelligence (AI) in assisting the creation process which is not experimented in the design fiction practice. Recently, AI-assisted tools are experimented in creativity activities such as drawing, music composition, story creation, and design ideation [29]. Built on these early attempts, we aim to investigate 'how AI assistance would influence non-designers' creation process of design fictions. The AI-assisted tool chosen in this study is Generative Pre-trained Transformer-2 (GPT-2, referred as the transformer), a powerful text-generation model based on unsupervised multitask learning. The transformer was trained with corpus from THUCNews and nlp_chinese_corpus, using Cloud TPU POD v3-256 to train 22w steps [40].And the transformer can produce Chinese texts based on a large-scale database of Chinese web texts and novels. The investigation looks at three aspects of the quality of fictions, participants' experiences with the AI agent, and most importantly, their ways of incorporating those texts in writing. Our investigation is based on writing workshops of design fictions with 130 Chinese youth studying different subjects in Southeast University. In the workshop, each student is required to produce a piece of fiction about a future world of automated work in 2040. The workshop is part of the university's liberal education on technologies. help Chinese youth develop and communicate their critical reflections on emerging technologies in the future of 2040. The 45 participants are divided into two groups. One group (N=20) used AI-assistance in their writing process and the other group (N=25) developed fictions entirely on themselves. The analysis focuses on the group that use AI assistance. After the workshop, we conduct survey and post-interviews with participants and also rate the fiction quality from both groups. Based on the data, we present our findings of the main values of AI assistance in the process of creating design fiction. For further design and use of AI assistance in creating design fiction and similar creative work, we propose two findings of enhancing interactional quality between humans and AI and prompt programming. Lastly, for the practitioners with similar interest in the participatory approach to future speculation, this study shows the potential of AI assistance in making the process more inclusive and diffuse with a larger number of audiences.

## 2 RELATED WORK

The investigation of AI assistance in design fiction creation is built on three bodies of work. Primarily positioned in design fiction practice, we firstly reflect the tools and strategies of facilitating participatory process of design fiction from the design community. Secondly, we examine some early attempts from the HCI field of applying AI assistance in creative activities. Built on the experiences and learnings, future opportunities and concerns are proposed. Thirdly, we review the ways of planning and studying creative writing processes from creative writing scholars.

### 2.1 Tools, Strategies in Creating Design Fictions with Non-experts

Design researchers have made great attempt to develop tools and processes to better craft design fictions for both design experts and non-experts. Recently, increasing efforts are on engaging non-experts in the participatory design fiction process [37] . The common context is the co-design workshop with a specific design investigation of a yet-to-exist technology in which non-designers such as future users or citizens, stakeholders, or company partners are invited. However, it is especially challenging for non-designers to imagine beyond their daily lived experiences. Therefore, in co-design workshops, a wide range of prompts are provided to feed or spark imagination, such as related articles, clips of TV shows or movies [7], questionable design concepts [58], and narratives of imaginary technologies [2]. Moreover, the provision of backdrop has been found useful for participants to take a leap from the present, enter the fictional world, and get sensitised with the futuristic ambiance. For instance, in Baumann et al.'s [4] co-design project of envisioning future neighbourhood with urban technologies, researchers carefully embedded design brief in 'what if' prompts for participants, like 'what if self-driving shuttles replaced privately owned vehicles?'. Another strategy is seen in Wu et al.'s [62] work which envisions a future of autonomous shipping in 2035. The co-design workshop with company partners was structured based on the storyline that experts and company developers 'travelled' to a year when the automated system of shipping had been implemented and influenced the society. In addition, Cheon & Su [12] applied the technique of 'autobiography' that invited participants to explain how the story was developed from the present.

When facilitation materials are various, participants' ways of contributing to the creation process are also different. For instance, participants co-compose their imagined futures [2], develop the half-finished storyline with role-playing [62], or give opinions to the written fictions by experts [1, 57, 58]. Lately, a more performative and interventionist approach is experimented [11]. Noortman et al. [43] gave participants a prototype of remote care device and asked them to play the role of a caregiver for dementia patients. Similarly in the topic of elderly care, Ng et al. [42] built a fictional online shopping page of a social robot for the Spring Festival of 2035 and asked users about their purchase decisions and further stories.

### 2.2 AI Assisted Tools in Creativity Support

The advancement in artificial intelligence (AI) enables AI to play an increasingly significant role in creative work such as drawing and music. The enquiries of applying AI in art and design activities and human-AI collaboration have been of the interest of the HCI community. However, the design for better human-AI interaction is full of challenges across the design process [63]. For instance, the unpredictability of AI's capability, outcome, and error makes designers difficult to ideate functions, plan use scenarios, or test prototypes. Despite of the challenges and difficulties, the rapid development in AI such as learning algorithm [46] or large-scale models [9] brings new possibilities that researchers cannot resist. In recent years, there are emerging early attempts that explore AI support in creativity such as creative writing [15, 35, 44], drawing [21, 47, 65], design ideation [30, 31, 34]. As Chung et al. [14] observe that the HCI field has developed more tools to support vision-oriented activities of idea generation rather than skill-improvement such as revision or implementation.

Studies have shown the two main contributions of AI-assisted tools in creativity support which are to inspire and to motivate. To inspire, however, researchers take a careful attitude towards keeping the autonomous role of human writers in the collaboration. Suggestions such as 'Machine in the loop' [15], 'Say Anything' [56] and 'Creative Help' [50] are provided so that when human get inspired they still have the final control over the final result. Osone et al. [44] are more cautious, who focus on improvement based on writer's written content and further increase their interest in creating more new ideas on their own. Apart from suggesting new ideas, information and knowledge [14], motivating writers is also regarded crucial. Therefore, the format of games is a common engagement strategy to consider, such as role-playing [44] and mystery game [35].

Among the successful early endeavours of AI assistance in creativity support, they share the sense of excitement on various new opportunities brought by more powerful AI in terms of reliability in coherent performance, predictability in output [14], or more human-alike interaction and behaviours [28, 45]. However, while most of the attention is paid to the improvement of machines skills [23], rare looks at user experiences related to interface or collaboration and the specification of intended user groups [14]. However, there is an exception that Hwang & Won [30] locate user experiences in teamworking in ideation. By closely studying the differences of robotic-sounding and a human-like conversation styles, the study indicates the important concerns, such as perceived image of the AI agent, and self-efficacy of human, for further research. The current stage of these exciting and new possibilities poses a good question to designers: which improvement or new possibility do we want to focus on?

### 2.3 Creative Writing: Exploring the Making of a Creative Mind in Laypersons

Creative writing studies provide this paper with insights regarding planning and studying the creative writing process. Creative writing is a creative behaviour in nature and the generation of creative thoughts can be synthesized as a four-stage process [59]: the first one being preparation, in which the topic or problem is investigated in all directions; the second one being incubation, in which the author is not consciously thinking about the problem; the third one being illumination, the appearance of the 'happy idea' together with the psychological events which immediately precedes and accompanied the appearance of ideas; the fourth stage being verification. And for short-story and novels, Crowley [16] gave out a four-stage process: 'the germ of the story', 'the period of more or less conscious meditation', 'the first draft', and 'the revision'. And for the more poetic way of expression, Wilson suggests five stages [61]: 'the selective perception of the environment'; 'the acquisition of technique the envisioning of combinations and distillation'; 'elucidation of the vision', and 'the end of the poem and its meaning to the poet'. Theories of the creative process of different genres helped us see the writing of design fiction as a situated practice that undergoes a series of cognitive and behavioural stops before coming into being.

To study and understand the creative writing process, as Emig [20] put it, traditionally there are three major types of data. The first type is the accounts concerning writers, such as a writer's description of or reflection on one's own method of writing and revision. For instance, 'author talk' is the dialogue or correspondence between a writer and 'a highly attuned respondent' such as an experienced editor or a fellow writer, and critical analysis of the evolution of the writing from sources and revisions. The second type is to directives and handbooks of creative writing by authors, editors of rhetoric and compositions. The third type is the research of 'the creative process' of writing among adolescents. Aiming at a triangulation approach, we formulate our research plan using the first and third type of data with multiple methods, such as surveys of participants' writing experience, interviews, and thematic analysis of writing products. We believe that the accounts about the writing process, made both by the author and may provide useful insights into the ideation of design concepts and storyline. The products of the writing process also speak to the use of AI-generated responses.

In creative writing, new technology and softwares, such as *Storybird* and *Write Your Journey*, have been used to assist with writing skill learning [13], but we have not seen a particular technology designated to develop writer's imagination and creativity. In this sense, we believe that our study may contribute to creative writing studies with regards to the breadth of research.

## 3 CASE, DATA

### 3.1 Background: Reflecting on Future Work

The creation workshop of design fictions was part of the course *English Technical Writing* at Southeast University. It was a course that is designed to foster technical writing skills and critical reflections on human and technology. The workshop required each participant to produce a piece of fiction about *future work* in 2040 with emerging technologies such as automation, big data and so on. We conducted the writing workshops in two formats–one with AI tool (referred as *the AI Group*, N=20) and one with no AI tool (referred as *the Non-AI Group*, N=25). All 46participants in both groups were second year college students, the majority of whom majoring in engineering. The non-AI group generated 25 design fictions with each participant completed one. The *AI Group* generated 25 design fictions with two participants voluntarily wrote three, one participant wrote two, and the rest participants wrote one. The *Non-AI Group* had 25participants who were enrolled in *English Technical Writing* course. As for *the AI Group*, because the use of AI tool required a Google account that was not publicly supported by the internet use policy in Chinese universities, we were able to recruit only 20 participants who privately used VPN to access Google. Participants in *the AI Group* were recruited through a recruitment post that was electronically distributed campus-wise. n addition, all the participants, both in the *Non-AI Group* and *AI Group*, were second year students who have been learning similar core curriculum courses since they entered college.

The AI-assisted tool of GPT-2 is an open-source artificial intelligence created by Open AI. The transformer in our study was trained with corpus from THUCNews and nlp_chinese_corpus [40]. This transformer was trained to predict the next word after users inputting original texts. In our workshop, we used the Chinese version of GPT-2 (Figure 1). All the design fictions generated in the workshop were written in Chinese.

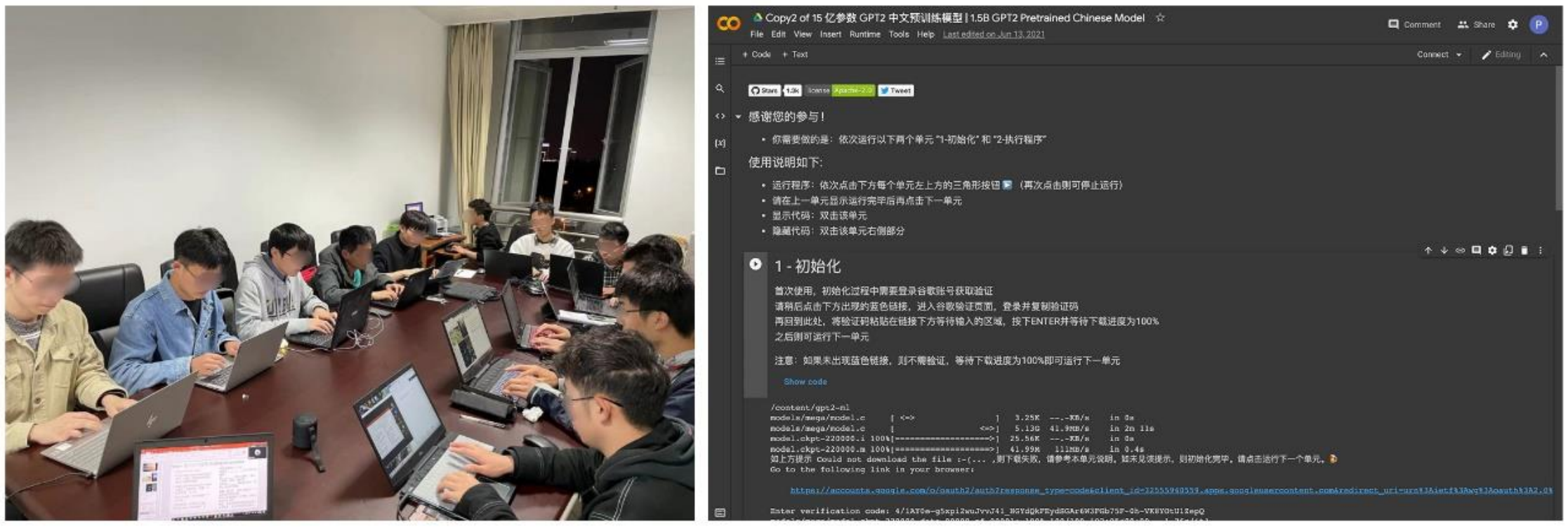

Figure 1: Writing workshop of *the AI Group* and a screenshot of the GPT-2 interface

The purpose of the design fiction workshops was two-folded. On the one hand, it served as an integrated part of the course, providing students with thought provoking learning materials as well as inspirational learning experience; on the other hand, the design fiction workshop aimed to bring about or foster awareness of the challenges faced by humans in the near future. Design fiction, treated as a special *genre*, was adopted as the *pedagogical tool* to help participants generate, develop, and communicate ideas and reflections into a narrative form.

### 3.2 The Creation Process: A Future World in 2040

The three-hour creation process had four steps of 'Entering the fictional setting', 'Setting a goal', 'Taking action' and 'Developing consequences' (Figure 2).

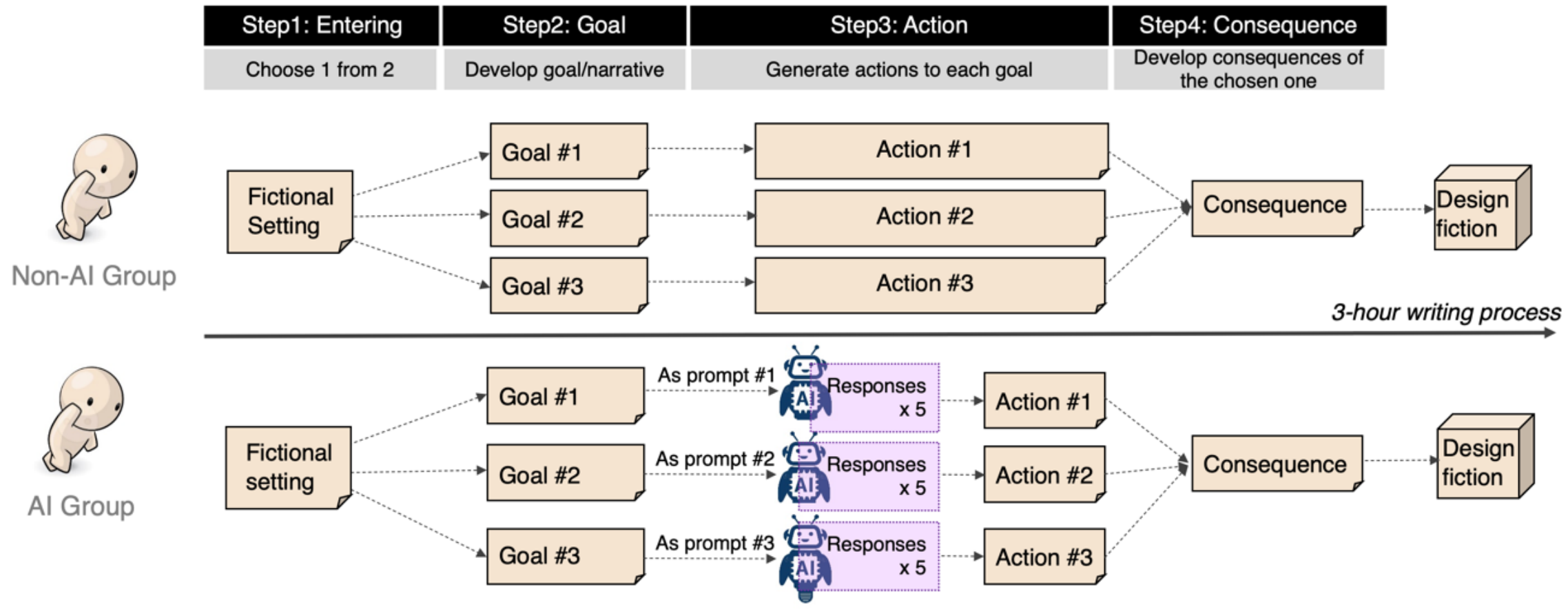

Figure 2: The creation process with four steps

In Step 1, to help participants to get familiar with the fictional setting of future work, we provided two future scenarios for participants to choose one to enter. The first fictional social setting *'The Age of AI Involution'* portrayed the over-heated competition between humans and AI in the work domain, developed the current controversial phenomenon of the involutionized competition especially in the Chinese IT industry. The second one *'Data generators as Workers'* described that people get income by generating and selling data from daily living, developed from the current discussion of 'private data as capital'.

> **Scenario 1/ The Age of AI Involution** (EN trans): In the year of 2040, AI and automation have replaced a significant portion of human work. While productivity has been improved to a certain extent, the involutionized competition has become more serious. Moreover, the competition has spread from the involution between humans to the ones between human and AI, even between AIs. For example, the human survivors of competition begin to compete with AI. Some automation technologies are developed not to increase productivity, just to win the involution.

> **Scenario 2/ Data Generators as Workers** (EN trans): In the year of 2040, AI and automation have replaced a significant portion of human work. The types of work have become diverse, not limited to the ones of traditional definitions. 'Living as work' has long been mainstreamed as a means of bringing in fractional income for many people. For example, as long as people live, they can generate data through a variety of behaviours in exchange for income. Even in sleep there is a chance to make money. Of course, the list price is not the same for different behaviours.

In Step 2, after choosing a social setting between Scenario 1 and 2, participants developed further narratives including a goal or a problem of the individual character. They were requested to develop three different versions. In Step 3, *the AI Group* and *Non-AI Group* had different activities. *The Non-AI Group* developed solutions or actions by themselves. While the AI Group input the result of Step 1 and 2 as the prompt *'In 2040, there is the fictional situation. I have a goal to achieve. Therefore, I …'*. Responding to the one prompt, the AI-assistant tool produced five responses and each response had a minimum of 5 sentences and a maximum of 8 sentences as we pre-set. However, the generated responses were very diverse in length. Based on AI-generated texts, participants from *the AI Group* were requested to develop one consistent storyline of 'goal-action' from each prompt. In Step 4, participants chose one out of three to develop societal and ethical consequences. And eventually each participant produced one piece of design fiction.

The two social settings provided in Step 1 were to facilitate participants to enter the future world with the provision of backdrop [4]. And the plot structure took the common fictional structure of 'overcoming a monster' [7] with both elements of 'goal/problem' and 'actions/solutions'. Lastly, in Step 4 'Consequence', participants were especially encouraged to develop critical results with 'friction' [22] to create 'a discursive space' that features the genre of design fiction [36]. In the end, we have 20 participants in the AI Group and 25 participants in the Non-AI group successfully completed the tasks , with 25 valid pieces of fictions from the former group and 25 pieces from the latter.

### 3.3 Post-workshop Enquiries

After the creation process, we did the post-workshop survey and interviews. To *the Non-AI Group*, we asked them about the experiences of the writing process. To *the AI Group*, we added more questions related to the AI tool, including their self-reflections on the usage of AI-assistance in their writing process, overall impression and user experiences of AI-assistance, and their perceived perception of AI's influence on ideation and writing, especially imagination. Moreover, we conducted one-on-one semi-structured interviews with each *AI Group* participant immediately after their creation process. In the interviews, we asked them to describe in detail what kinds of AI-generated texts they found attractive, useful and how they used these texts in their writing.

Apart from the enquiry of process and experiences, we evaluated the quality of design fictions produced at both workshops. We developed a set of quality criterion and recruited five human raters to evaluate the design fictions. 25 pieces of design fictions were randomly selected from *the Non-AI Group* and 25 were produced by *the AI Group.*

Lastly, we used the criterion of five dimensions to evaluate the fiction quality (Table 1). The criterion was developed with the consideration from the practices of both design fiction and creative writing. As a *genre* of writing, design fiction shall meet the basic requirement of *writing quality* in clarity and coherence (Table 1). The other four dimensions were specially made for the creation of a piece of fictional narrative about the future world in 2040 with critical thinking of technology. Therefore, we expected the design fiction to be *provocative* to open up discursive space, *plausible* for made-believe, *creative* in the design sense, and *concrete* in scene building. It is worthy to note that we agree with Baumer et al.'s [5] opinion on the impossibility of making one single standard evaluation of design fiction and the hesitation on the necessity of the evaluation. Nevertheless, we used this evaluation criterion to fulfil the research purpose of understanding the features that AI assistance might have an impact on.

The evaluation criterion was developed through three stages: We firstly developed a draft version; secondly, we have it reviewed by three experts from relevant fields of design fiction and psychometrics; thirdly, after one round of discussion with the experts, we revised and finalized the criteria (Table 1). For the measurement, we invited five raters, all of whom were graduate students in design studies or literary studies. Two raters had profound knowledge and practical experience in design fiction, and the other three raters were majoring in literature study and were also avid raters of science fiction. The five raters all participated in a training session, in which they received instruction on rating criteria and process, and completed three rounds of testing and discussion to ensure inter-rater reliability. Later on, raters measured design fictions on 5-point Likert scales.

Table 1: Rating criteria of design fictions

| Criteria | Description |
| --- | --- |
| 1. Writing quality | Overall, the writing is clear and easy to understand for a reader.<br>•The flow of writing is easy to follow. Or,<br>•The lexical and structural choices are appropriate. Or,<br>•It does not have unintended ambiguity or confusion. |
| 2. Provocativeness | Overall, the text could trigger meaningful conversations or new thoughts.<br>•It could trigger reflections on technology, taken-for-granted assumptions, or established structures. Or,<br>•It could reveal certain conflicts or tension. Or,<br>•It could trigger a new perspective to look at emerging technology or other issues<br>•This text could inspire new ideas in the audience. |
| 3. Plausibility | Overall, this fiction is self-consistent in light of its own fictional setting.<br>•Some connection can be found between this fiction and certain trends in reality. Or,<br>•The fictional technology, objects, and events are not completely magical and can make the audience believe that they are logical in the overall narrative. |
| 4. Creativity | •Strong imagination and creativity. Or,<br>•It has surprising, unexpected elements |
| 5, Concreteness | •This text has details that can help the audience to get immersed in the narrative. Or,<br>•The text describes some concrete, tangible event, conflict, or tension (stemmed from new technology), rather than only speaking of abstract concepts or speculations. |

### 3.4 Data Analysis

The collected data included both quantitative and qualitative data. The quantitative data was the rating result of the quality of design fictions, 25 pieces produced from *the AI Group* and 25 pieces randomly selected out of the 109 pieces from *the Non-AI Group*. We measured each participant's scores on the five dimensions and the total score, and we also tested the statistical significance of the differences between the two groups. The aim of the analysis was to assess whether the AI tool had had an impact on the quality of design fictions and how. The focus of the analysis is not on the comparative assessment of the performance difference between human and AI agent. Rather, we looked at which aspects of AI assistance would have more impact and which not.

The analysis of the qualitative data aimed to investigate user experiences of human-AI interaction in the writing process and the usage strategies. To achieve that, we used open coding through a thematic analysis approach [8] to identify themes without pre-set categories. Regarding the first part of user experience, the analysis closely looks at the interactional quality of human writers and the AI-assistant tool and their impression of AI-generated texts. Based on that, we concluded with the metaphor of 'fun dancing with drunk agent'. And the second part of usage strategies, we analysed what kinds of AI-generated texts were chosen, how they were used and how they helped human writers in creation. In the end, we argue for two main usage strategies of applying AI-generated texts.

## 4 RESULTS

### 4.1 Design fictions produced from *the AI Group* and *Non-AI Group*

The design fictions from the AI Group and Non-AI Group show distinctive features respectively. Figure 3 contains two word-cloud figures displaying details and contrast of key words in each group. The word clouds are generated through a NLP program, and the size represents the frequency of a word, with larger size equalizing a higher frequency. Since the design fictions were written in Chinese language, the words in the figure are a literal translation.

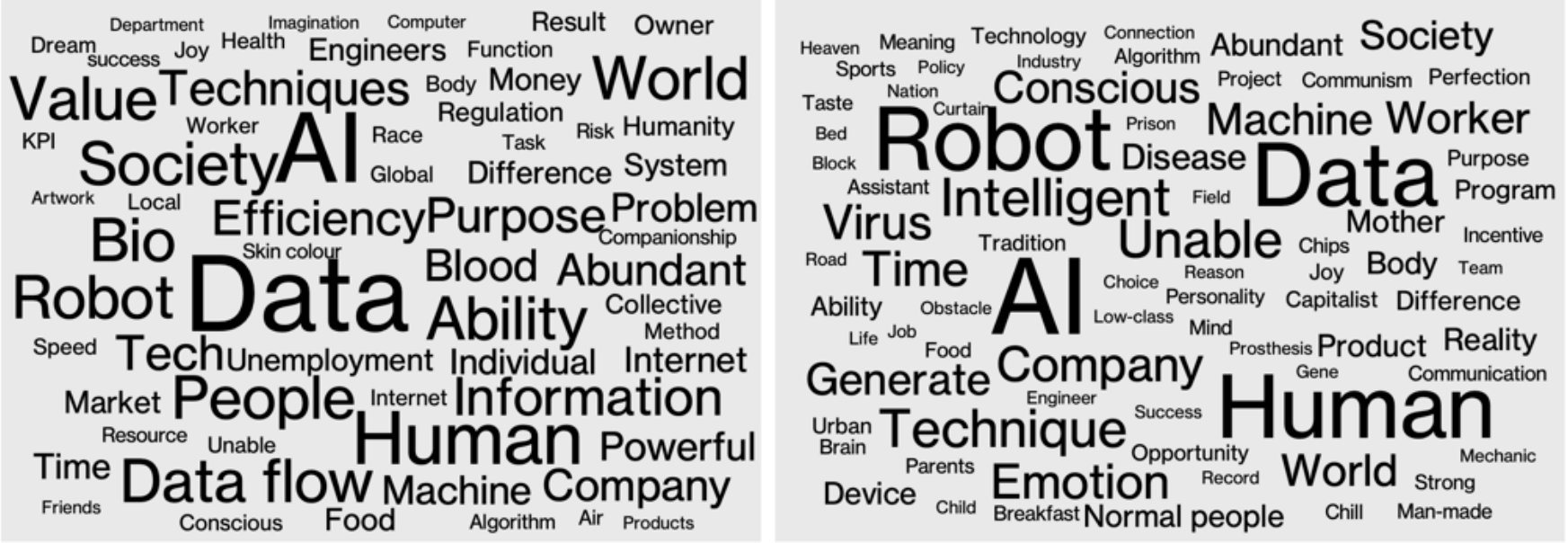

Figure 3: the two word-clouds that indicate the themes or topics being covered from the 50 pieces of design fictions; left (from *the AI Group*) and right (*the Non-AI Group*)

Among the 25 pieces of design fictions produced from *the AI Group*, two pieces from participants #7 and #9 (AP7, AP9) did not use texts or ideas from AI-assistance, and 23 pieces had AI-generated texts integrated in one way and another. Here, we give three pieces of design fiction to illustrate how AI-generated texts were integrated with the writing ranging from the large, middle to small extent (Table 2). The design fiction (from AP1) to a large extent was comprised of 73,7% of texts (314/426 words) that were directly taken from three AI responses. The one in the middle (from AP5) had 31,3% words (102/326 words) from AI. And the smallest extent, the fiction (from AP20) used the word 'teacher' as an inspirational concept in creation.

Table 2: Three examples of using AI-generated texts in writing: human texts (Yellow), AI-relevant texts (Purple)

| **Large extent** (AP1) | **Middle extent** (AP5) | **Small extent** (AP20) |
|---|---|---|
| 73.7% (314/426 Chinese words) | 31.3% (102/326 Chinese words) | 1% (2/311 Chinese words) |
| 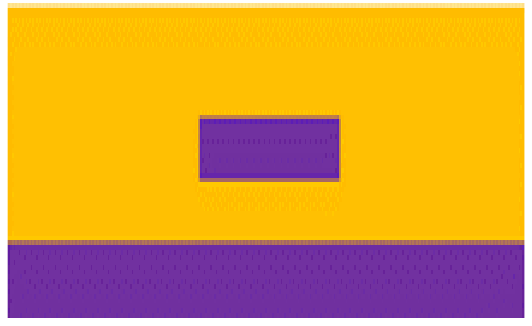 | 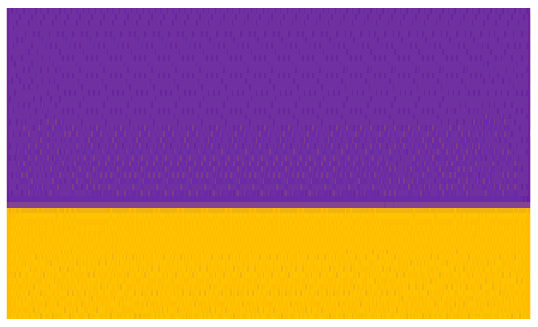 | 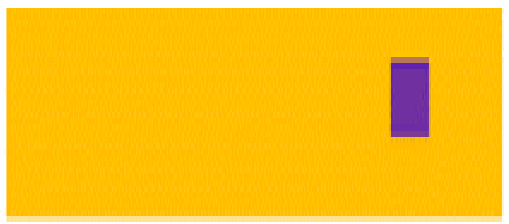 |
| Summary in English<br>In 2040, I earn money by selling data of my appearance. To achieve this goal, I analyze the definition of beauty in China.(...). I build a website to identify beautiful women through photos relying on AI tech. (details on how). In addition, I manage to sell data from other beautiful people. | Summary in English<br>In 2040, data becomes the currency, and I am a hacker who trades data for profit. So I formed a "Harvard Data Group". *(details about the Group).* Gradually, the Group has collected adequate data to do accurate prediction. | Summary in English<br>In 2040, AI technology has its strength and weakness in art creation. *(details).* Later, a new professional called ''AI teacher' is emerging to teach AI in art. |

## 4.2 Rating Result of Fiction Quality

To evaluate the quality of design fiction, we provide general descriptive statistics. Table 3 provides descriptive statistics of all quantitative measures. The results showed that *the Non-AI Group* scored slightly higher than *the AI Group* on all of the five aspects as well as the total score. To be specific, the AI agent did not seem to have a positive impact on the overall writing quality of the design fictions.

Table 3: Rating scores of the fictions from AI Group and Non-AI Group

| Dimensions | AI Group | | Non-AI Group | |
|---|---|---|---|---|
| | *M* | *S.D.* | *M* | *S.D.* |
| Writing quality | 2.91 | 1.15 | 3.51 | 1.12 |
| Provocativeness | 3.31 | 1.15 | 3.57 | 1.19 |
| Plausibility | 2.88 | 1.20 | 3.34 | 1.21 |
| Creativity | 3.07 | 1.09 | 3.24 | 1.14 |
| Concreteness | 2.83 | 1.08 | 3.36 | 1.16 |
| Total score | 74.88 | | 85.08 | |

Note: Total score=Rater1+Rater2+Rater3+Rater4+Rater5

## 4.3 Participants' Self-Reflection on Imagination

To investigate AI assistance's impact on participants' imagination, we developed a survey with 9 questions adopted from the Four-Factor Imagination Scale (FFIS) [64] which focused on characteristics of the imaginative process and individual differences such as emotional valence and directedness of imagination. Shown in Table 4, among the positive dimensions, four parts were mostly agreed by participants, which were *P3 AI made their imagination richer and more complicated* (M=3.92), P4 *better visualize a setting in my mind* (M=3.68) and *P2 provided vocabulary hint* (M=3.88) and *P1 details* (M=3.44). Participants tend to disagree with the statement that *N1 triggering negative emotions* (M=1.92), *N3 making people lose directions and goals* (m=2.12), and *N4 making imagination depressed* (m=2.16). The result showed that participants self-projected the positive influence of the transformer in enhancing their capability and experiences related to imagination.

Table 4: Participants' feedback on AI-assistance's impact on imagination

| Item Label | Item | $M$ | $S.D.$ |
|---|---|---|---|
| P1 | AI-assistance provides rich details to my imagination | 3.44 | 0.96 |
| P2 | AI-assistance provides more hints to my imagination on the vocabulary level. | 3.88 | 0.67 |
| N1 | AI-assistance makes me have negative emotions. | 1.92 | 0.76 |
| P3 | AI-assistance makes my imagination richer and more complicated | 3.92 | 0.64 |
| P4 | AI-assistance helps me better visualize a setting in my mind. | 3.68 | 0.85 |
| N2 | AI-assistance makes me feel pessimistic. | 2.72 | 1.02 |
| P5 | AI-assistance gives me clearer directions and goals. | 3.48 | 0.82 |
| N3 | AI-assistance makes me lose directions and goals. | 2.12 | 0.67 |
| N4 | AI-assistance makes my imagination more depressed. | 2.16 | 0.94 |

## 5 INTERACTING AND COLLABORATING WIH A 'DRUNK' AGENT

This section has two parts. Firstly, we describe the joyful interaction that participants developed with AI-assistance which was perceived as 'a drunk, knowledgeable and fun agent'. Secondly, we present two ways in which the transformer helped human writers in their creation process of design fictions.

### 5.1 Joyful Interaction with the 'Drunk' Agent

Overall, human writers perceived the transformer as 'a drunk, and yet knowledgeable and fun agent', and their interaction with AI-assistance enjoyable. In answering the question 'list three keywords to describe AI-generated content', 20 participants provided 55 keywords (Table 5). They are divided in four categories of creative (N=25), inconsistent (N=17), diverse (N=7) and relevant (N=6). Below are three quotes that expressed the mixed feeling of AI-generated texts that were perceived as creative and inconsistent:

> 'Not knowing exactly what it talks about, but interesting and cute!' (AP17)
>
> 'AI's texts are more interesting and richer than I expected. But, often, they have nothing to do with what I wrote.' (AP15)
>
> 'AI has presented me unconstrained imagination. And it covers a good command of knowledge of all aspects. However, most of the stories cannot be called 'a story'. (They are) messy and incomplete! Sometimes, the next sentence is not even related to the previous one.' (AP10)

Table 5: Participants' expressed experiences of AI-generated content

| Creative | N=25 | Inconsistent | N=17 | Diverse | N=7 | Relevant | N=6 |
|---|---|---|---|---|---|---|---|
| Novel | 7 | Inconsistent | 8 | Rich | 4 | Relevant | 3 |
| Unexpected | 6 | Illogical | 6 | Uneven | 1 | Rational | 1 |
| Interesting | 6 | Random | 4 | Abundant | 1 | Logical | 1 |
| Provocative | 3 | Irrelevant | 2 | Complicate | 1 | Natural | 1 |
| Strange | 1 | Confusing | 1 | | | | |

| | |
|---|---|
| Funny | 1 |
| Extreme | 1 |

When reflecting on their favourite part in the whole writing process, the highlighted experience was the interaction with AI-assistance. 18 out of 20 participants mentioned the experience related to the usage of the transformer. 10 participants regarded reading AI-generated texts as the most fun. The joy was mostly brought by the creative, unexpected and diverse content (‘*I feel like being inspired like firework by these weird ideas of all kinds’).* Even one participant found fascinated to examine the distance between AI and humans: *‘It is fun to see what new ideas AI can provide that I cannot.’* And 8 participants considered the next step of writing continuation based on AI-generated texts exciting, in which they found a strong sense of adventure and achievement in the collaboration with the interesting ideas provided by AI assistance (such as *‘the best part is I can start from a novel seed from the texts and use my imagination to complete the story.’*).

An interesting finding is that the more creative process (the AI-incorporated process) yielded less creative products. We want to point out that our finding is in accordance with Mumford, Connelly, and Gaddis [66] finding that *idea evaluation* is negatively correlated with the performance of creativity. We argue that the *idea evaluation* in this study took place during participants' interaction with AI tool. Once the writers got responses from AI, they were actively reading, thinking about, and digesting AI's ideas, which resembled the *idea evaluation* process. Thus, the active digesting, or "idea evaluation", lead to a less impressive performance. The reason why idea evaluation leads to poor performance is unclear. We call for more research on this topic.

### 5.2 Two Ways of AI Incorporated into the Creation of Design Fictions

Next, we describe two usage strategies of AI-generated texts in the creation of design fictions. With the first strategy, human writers took the concept from AI-assistance as the direction for narrative development. It was often a new direction which was different from what they expected previously. They preferred the unexpected ideas to expand their mind. With the second strategy, participants used their own idea and used AI-generated texts to enrich and concretise scenes.

#### *5.2.1 Suggesting Creative Directions with the Unexpected*

As reported, 14 participants took the ideas from AI-assistance directly as the direction of fiction development. They were attracted by the content that was unpredictable, dramatic or even contradictory. Furthermore, they took the idea from AI-assistance as either the basic storyline or the design direction to further generate new design concepts of products and services. Participants who adapted this strategy were willing to start a new journey in which they were never able to come up by themselves. Here are two examples of the inspiringly odd turn:

> -Human’s prompt: ‘In 2040, I feel ashamed to have an AI dog as the pet. I wish to get a real dog.’ AI’s response: ‘Therefore, I get a new dog every month’. (from AP7, unselected fiction)

> -Human’s prompt: ‘In 2040, I wish AI to do all housework to bring convenience to my life.’ AI’s response: ‘Therefore, I start a romance with an AI robot.’ (from AP4)

In addition, we present one completed case of fiction writing to show how the human writer composed design fiction inspired by the unexpected solution from AI (Table 6). The participant (AP3) wanted to develop actions to have more sweet dreams. Among the five AI responses she quickly noticed the response '4/5 *Therefore, I collect dreams from others'*. The author mentioned that she was expecting some technological devices to generate more dreams and the idea of 'from people' appeared unexpected but sweet. Therefore, she decided to shift the direction. The comment this fiction received from raters is '*a great creative idea, also warm and optimistic. Different from other stories on confrontation wars between human and AI'*.

Table 6: The example of fiction 'dream generation'

| Design fiction (AP3): 'Dream Generation' |
|---|
| Human Writer: Prompt (Step 1+2: Entering + Goal)<br>In 2040, dreaming in sleep can bring income. I have been selling my sweet dreams to support my humble living. However, I wish to have more sweet dreams to bring more income. Therefore, I … |
| AI: 5 responses (Step 3.1)<br>1/5 Therefore, I produced the biggest dream coded 1108. (…)<br>2/5 Therefore, I learn English everyday before sleep. (…)<br>3/5 Therefore, I called more banks to negotiate. (…)<br>4/5 Therefore, I collect dreams from others. (…)<br>5/5 Therefore, I try hard to sleep more. (…) |
| Human writer (Step 3.2):<br>Therefore, I talk to happy and kind people everyday to gain more resources and mood for nurturing my dreams. In the beginning my purpose is to earn more money. However, the more I talk, the more joy I get from these conversations. (…). Data economy does not have to be an explorative game. It has the opportunity to bring happiness. |

### *5.2.2 Enriching and Concretizing World Building with the Diverse and Beyond*

In the second strategy, AI-assistance is used to enrich and concretise the building of fictional scenes. With this usage strategy, the human writer was mostly responsible for the basic idea of the fiction and just borrowed or reframed the keywords or phrases from AI-assistance to enrich the fiction. And we observed that such inspiration happened quite random and personal that a newly added element can be a character, a philosophical comment, a technical term, or a piece of concrete imagery. For instance, a participant (AP17) mentioned that his strategy was intentionally seeking unfamiliar terms or scenes such as '*holy shrine'* that he would never come up based on his own life experiences. And some other participants found the concrete names of a new technology or company very useful. They directly quoted to name their newly created product or institute to make the fictiveness more concrete and vivid.

Below we show two fictions respectively from *the AI Group* and *the Non-AI Group* addressing the similar topic of unemployed people selling data to live (Table 7). Both writers started with an initial piece containing limited information. When AI-assistance jumped in, we see a surge of new ideas and concepts of all sorts. As a result, fiction #16 appeared jumpy and contains more elements both in quantity and variety, with a widen plot and a radical turn. While fiction #8 was more consistent in logic and elements. The AI-generated responses were indeed 'jumpy', and 'all over the place,' but they helped writers literally think outside the box and produce much diverse and rich ideas.

Table 7: Summary of the two fictions

| #16 from *the AI Group (from AP16)* | #8 from *the Non-AI Group* |
|---|---|
| Step1+2: Entering + Goal | |
| In 2040, AI has replaced most human jobs. I am a useless motorcycle repairman at 30YO. Although I keep honing my | In 2040, AI replaces many basic tasks in human society. And the phenomenon of digitization is serious. People |
| skills and abilities, no company wants me. Everything in my life seems to be against me. I need new opportunities and a motive to live. Therefore, I… | living in cities have become tools for providing data for AI. They hope to get rid of the fate of being a tool and live as a real person. |
| Step 3: Action | |
| The AI's response that inspired human writer:<br>I made up my mind to join the Argo industry. Argo is a data collection system that has existed for a long time. | A solution created by human writer:<br>I invented a portable device that can instantly eliminate the data generated by various behaviours, so that users cannot be recognized by AI. |
| A summary of the design fiction | |
| In 2040, AI has largely replaced human jobs.<br>I was a programming engineer but lost my job at 40.<br>I started to look for new hiring opportunities online. I know that the internet AIs are growing as they are getting nutrients----the data generated through every user's online activities. However, I have no choice but to use the internet.<br>Luckily I accidentally got into a forum in the Dark Web. It is from this forum that I heard Argo Associate for the first time. Argo is …… *[detailed description of Argo.]*<br>One day, surprisingly Argo called me. *[a long dialogue between me and Argo staff--- to involve me with all kinds of deal]*<br>I am inclined to accept Argo's offer. | In 2040, AI technology has replaced a considerable part of human work.<br>I am an engineer and lately I am laid off.<br>I have to sell the data of my everyday life to make a living.<br>Later I invent a device that can erase the data I generate. Sadly, a lot of people use this device to commit all kinds of crime.<br>AI finds out my device and soon I got arrested and put into prison. AI deactivates my device.<br>In prison, I start to reflect and believe that AI cannot overtake human because human is way more unpredictable. |

## 6 INTERACTIONAL QUALITY AND PROMPT PROGRAMING IN HUMAN-AI INTERACTION

In this session we present two findings based on the use behaviours from this study. With each finding, we propose two implications for further research of how to better design for and use AI assistance in design fiction and similar creative work.

### 6.1 Enhancing Interactional Quality Between Human-AI Rather Than Improvement of AI

This section discusses the interactional quality between human and AI. We begin with the finding of the attitudes that human writers developed towards AI-assistance, which is the respect and curiosity to the beyond and unfamiliar, and at the same time, held a strong sense of control. On the one hand, although AI-assistance appeared a 'drunk' agent, surprisingly human writers did not develop a refusal or resistant attitude or found it useless. Instead, they held an open, respectable and curious attitude towards AI. They could easily spot the unknown or unfamiliar information, moreover, tried hard to find relevance from the discursive (but interesting) generations and to seek inspiration from the unexpected. Moreover, they were willing to embrace the novel and unexpected perspectives, push their imagination, and step out of their comfort zone. On the other hand, they had a cautious distance from AI and presented a strong sense of control. Although some had high expectation on the intelligence of AI before using the AI assistant tool (e.g., that the AI assistance might automate their process of writing), they soon realized most of the AI-generated texts were intriguing but not meant to be immediately usable.

In all, we observe that human trusted AI assistance, regardless of the current technical limitations, to have the potential to be helpful. It shows a good balance between co-creation and autonomy in human-AI collaboration. Based on this finding, we propose future design should not focus on the weakness of AI performance solely or the improvement of machine skills. Instead, the focus should be on the relational quality of the collaboration of both sides. For instance, despite the incoherency of AI-generated texts, we do not suggest the direct solution of increasing the coherency. If AI-generated texts are coherent and flawless, human writers might be directly quoting rather than mobilizing their own reflectiveness and creativity to build further upon the inspiration from the AI-generated texts. Is this the kind of human-AI interaction that we really desire in the context of design fiction writing? In any part of writing design fiction that is meant for developing writer's critical and speculative thinking, AI should assist and provoke, or even pose meaningful seams, rather than replacing human or fully automating their creative processes.

### 6.2 Prompt Programming '*Therefore, I*'

The next learning is related to prompt programming, which has been an emerging topic in human-AI interaction [49, 55]. Based on several experiments before the workshop, we developed the prompt *'I have a goal to achieve. Therefore, I'*. After trying it out with 20 participants, we propose two features of the prompt that would be valuable for other design researchers. They are the clear logic of 'goal-action' and first person '*I*'.

We developed the clear logical structure of 'goal-action' as the turn taking between the human-AI collaboration. Moreover, participants were particularly asked to add the logical transitional phrase '*Therefore, I'* at the end of the prompt. Such prompt could make use of the written content from the previous steps as the crucial condition for AI-assistance to generate further narratives. The specific use of '*I*' was to help focus on generating or paying attention to the subject narrative of first person. This prompt programming enabled humans to write with a clear expectation, to quickly filter out irrelevant information, and moreover, to find the subtle balance between what is irrelevantly odd and what is inspiringly odd. However, to achieve such, the responsibility is not on AI as the generator but on human as the picker. Therefore, we propose that a clear logic and objective at the human's side can keep the good balance between the focused and disperse and further achieve an efficient and agile inspiration.

Our exploration of prompt programming is still initial. How to design or evaluate the efficiency of prompt to reach expected outcomes is increasingly important, yet under-addressed in AI-supported creativity and interaction (as well as in other domains related to generative AI and HCI). Implied by the prompt engineering of '*Therefore, I'* in our study, we are expecting new means of coding and programming with large generative AI models using natural language instead of computer language in the future.

## 7 CONCLUSION

This study is an exploratory attempt of investigating the influences, value and potential of AI assistance in creating design fiction by non-designers. Study results show the AI-assisted tool has little impact on the fiction quality, and the main contribution refers to user experience in the creation process especially the divergent part. Firstly, it brings joyful and excited experiences to the writing process, sharing the same result to previous work such as Clark et al.'s [15]. More importantly, the study has demonstrated the value of AI-assistance in the specific creation process of design fiction. Related to creative writing, design fiction is a special genre that needs to construct a fictional technological world that is yet-to-exist. And related to ideation, design fiction requires critical thinking of emerging technologies instead of the merely problem-solving approach. Moreover, the ideation is complex. More than ideating new products or services, it also ideates new interaction and experiences, and also fictional value [3]. Our study shows that AI-assistance can help non-experts by suggesting unexpected and yet creative solutions to pave for idea generation and providing diverse elements in diverging, enriching and concretizing the fictional scene. It confirms the finding from previous work that argues the unpredictability of AI performance is not flaw but would benefit idea generation [17, 25, 33]. Furthermore, we suggest two findings for future design of AI-assistance in creating design fiction and similar creative activities. Firstly, we suggest that the focus shall be on the enhancement of relational quality of human-AI interaction rather than improving machine skills. Secondly, we propose two suggestions in prompt programming. However, our study only involved two conditions of using AI assistance and not. We suggest further research including the third type of process that uses non-AI assistant tools for creative writing and ideation, in order to produce more insights on the role of AI-assistance in supporting creation of design fiction.

Next, we are open to two future enquiries. The first enquiry reflects on the perspective of creative writing, with which we argue that the disparity (see Table 3) might evolve from the different treatments of the creative process. The treatments that the non-AI group receive – *the setting up of social scenario* and a given *plot structure*– is what researchers call a *language-based approach* in creative writing studies [53]. It is systematic and algorithmic in nature; however, the AI group, compared with their non AI counterpart, receive one more treatment– *the AI intervention*, which can be considered a *reference-based approach* that is "pluralist and multiplicitous" and open to various possibilities [53]. This study suggests that more research is needed to explore a more organic merge of creativity-generation methods. The second enquiry looks at the bigger picture of the conversation between the human writer and the large database on which the transformer has been trained. We know that any trained transformer is not neutral. The Chinese version of the transformer used in this study is fine-tuned on 1.5 billion words of unstructured corpora of Chinese internet texts. When the AI tool produces responses to the prompt, it delivers the tip of the iceberg of the discourses from the data sources, which can be an embodiment of normative beliefs and concerns from the real-world living experience. For instance, *anxiety* was manifested when AI-generated texts were the responses to the prompts that included words like 'appearance anxiety', 'involution in IT working culture' and 'AI replacing human labours'. These are the heated topics widely discussed on Chinese internet. Participants could read the anxiety, which are embodied normative anxieties of desires from Chinese online communities about the past, present, and future of work. We would call for future research looking at how the embodied normative stances of AI tools will influence people in developing future narratives and developing opinions on technological futures. It is especially worthy of investigating when our participants reported that they would be more easily attracted and intrigued by the texts that were unexpected, novel, extreme and dramatic.

Lastly, in contributing to design fiction practice, this study is one of the pioneering attempts that apply design fiction in the context of technology and education [26, 48]. This study indicates the potential of applying design fiction to empower the mass with the critical discussion 'what kinds of technological futures do we want'. To achieve that, AI-assisted tools can make the process more diffuse and inclusive with larger audience and requires less work from facilitators. To engage with non-designers with no design skills, this study starts with the narrative power of design fiction [2]. Although the fictions in this study did not use diegetic prototype [6] and some might not appear 'designerly' according to design professionals, Design fiction's ability to democratize speculative and critical thinking shall not be undermined. Given its flexibility in facilitation and its affordance in fostering critical reflection, we believe that design fiction can have a place in ethics education and continues to uphold the advocacy of equity, justice, and access.

## Acknowledgement

This research project is supported by Foreign Language Teaching and Research Council, China Association of Higher Education (grant #21WYJYYB04). We express our gratitude to the experts in design fiction and technical writing in helping with making and commenting on the rating dimensions: Zhiyong Fu, Li Zhang, Yi Yang, Eric P.S. Baumer, and Marie Louise Juul Søndergaard.